\documentclass{PoS}
\usepackage{amsmath}

\title{Charm semileptonic decays at LHCb}

\ShortTitle{Charm semileptonic decays at LHCb}

\author{\speaker{Adam C. S. Davis}%
       \thanks{On behalf of the LHCb Collaboration}\\
      Center for High Energy Physics\\
      Tsinghua University\\
      Beijing, P. R. China 100084\\
      E-mail: \email{adavis@cern.ch}}

\abstract{In these proceedings, we explore the possible reach of the LHCb dataset in the area of charm semileptonic decays. Specifically, we give prospects for the measurement of $|V_{cs}|/|V_{cd}|$ using $\mathcal{B}(D^0\to K^-\mu^+\nu_\mu)/\mathcal{B}(D^0\to\pi^-\mu^+\nu_\mu)$ with Run I data. Preliminary projections show that the LHCb Run I dataset would give a relative statistical uncertainty of $\sim 0.2\%$ on this ratio. We also motivate the search for lepton non-universality in the charm sector.}

\FullConference{ 9th International Workshop on the CKM Unitarity Triangle\\
		 28  November - 3 December 2016\\
		 Tata Institute for Fundamental Research (TIFR), Mumbai, India}

\begin{document}

\section{Introduction}
The LHCb detector at CERN\cite{Aaij:2014jba} is a single arm forward spectrometer designed for precision 
studies of $b$ and $c$ hadrons in collisions at the Large Hadron Collider (LHC).
During Run I of the LHC, the LHCb detector collected approximately 3 fb$^{-1}$ of data 
at $\sqrt{s}=$7 and 8 TeV, and in Run II, approximately 2 fb$^{-1}$ to date. 
Between 2011 and 2016, LHCb has reconstructed approximately 1.8 billion charm hadrons. 
In these proceedings, we explore the physics reach of the LHCb dataset with 
respect to semileptonic $D$ meson decay.

These proceedings are broken into four sections: First, a brief review of the formalism 
of semileptonic decays in charm, including the relevant differential decay rate. 
Second, we explore the experimental challenges present for LHCb and address specific concerns for 
neutrino reconstruction and $q^2$ resolution, where $q$ is the momentum transfer to the lepton and neutrino. 
Third, we present sensitivity 
estimates for the measurement of $|V_{cs}|/|V_{cd}|$ 
or form-factors from the ratio of decays $D^0\to K^-\mu^+\nu_\mu$ and $D^0\to \pi^-\mu^+\nu_\mu$\footnote{Unless explicitly
stated otherwise, charge-conjugate decays are implied}.
Finally we present the motivation for the search for lepton non-universality in the charm system.
\section{Theoretical Overview}
\subsection{Differential decay rates}
The differential decay rate of the $D^0$ meson to a pseudoscalar final state $P$ via semileptonic decay
at leading order can be written in the following well-known form\cite{Aoki:2016frl}
\begin{equation}
\label{eq1}
\begin{split}
	\frac{d\Gamma(D^0\to P^- \ell^+ \nu_\ell)}{dq^2} = |V_{cQ}|^2 & \frac{G_F^2}{24\pi^3}\frac{(q^2-m_\ell^2)^2\sqrt{E_P^2-m_P^2}}{q^4m_{D^0}^2}\\ &\times \left[\left(1+\frac{m_\ell^2}{2q^2}\right)m_{D^0} (E_P^2-m_{P}^2)|f_+(q^2)|^2 + \frac{3m_\ell^2}{8q^2}(m_{D^0}^2-m_P^2)^2 |f_0(q^2)|^2\right].
\end{split}
\end{equation}
Here, $Q$ represents the outgoing quark from the weak vertex, the terms $f_+(q^2)$ 
and $f_0(q^2)$ are the vector and scalar form factors, respectively, 
used to parameterize the hadronic current.

The differential decay rate measured at experiments can be broken down into two pieces of interest. 
First, on the right hand side of Equation~\ref{eq1} starts with CKM factors $V_{cQ}$, which are 
of interest in testing unitarity of the CKM matrix. 
Second the right hand side depends on the vector form factor $f_+(q^2)$ and 
scalar form factor $f_0(q^2)$. Measurements of the form factor dependence provides useful input to 
Lattice QCD calculations. The interesting measurements which could be 
made by LHCb are:
\begin{enumerate}
	\item Measure the differential branching fractions for $D^0\to h^- \mu^+ \nu_\mu$ as a function of $q^2$. This helps constrain the quantity $|f_{+}(q^2)|^2 |V_{cQ}|^2$ and $|f_{0}(q^2)|^2 |V_{cQ}|^2$
	\item Using external constraints on the form factors from Lattice QCD, measure $|V_{cQ}|$ directly relative to a normalization channel
	\item Using external constraints on $|V_{cQ}|$, measure the form factor dependence, relative to a normalization channel
	\item Test lepton universality using decays which differ only by the lepton in the final state
\end{enumerate}
\section{Experimental Challenges}
\begin{figure}[htbp]
\begin{center}
\includegraphics[width=0.6\textwidth]{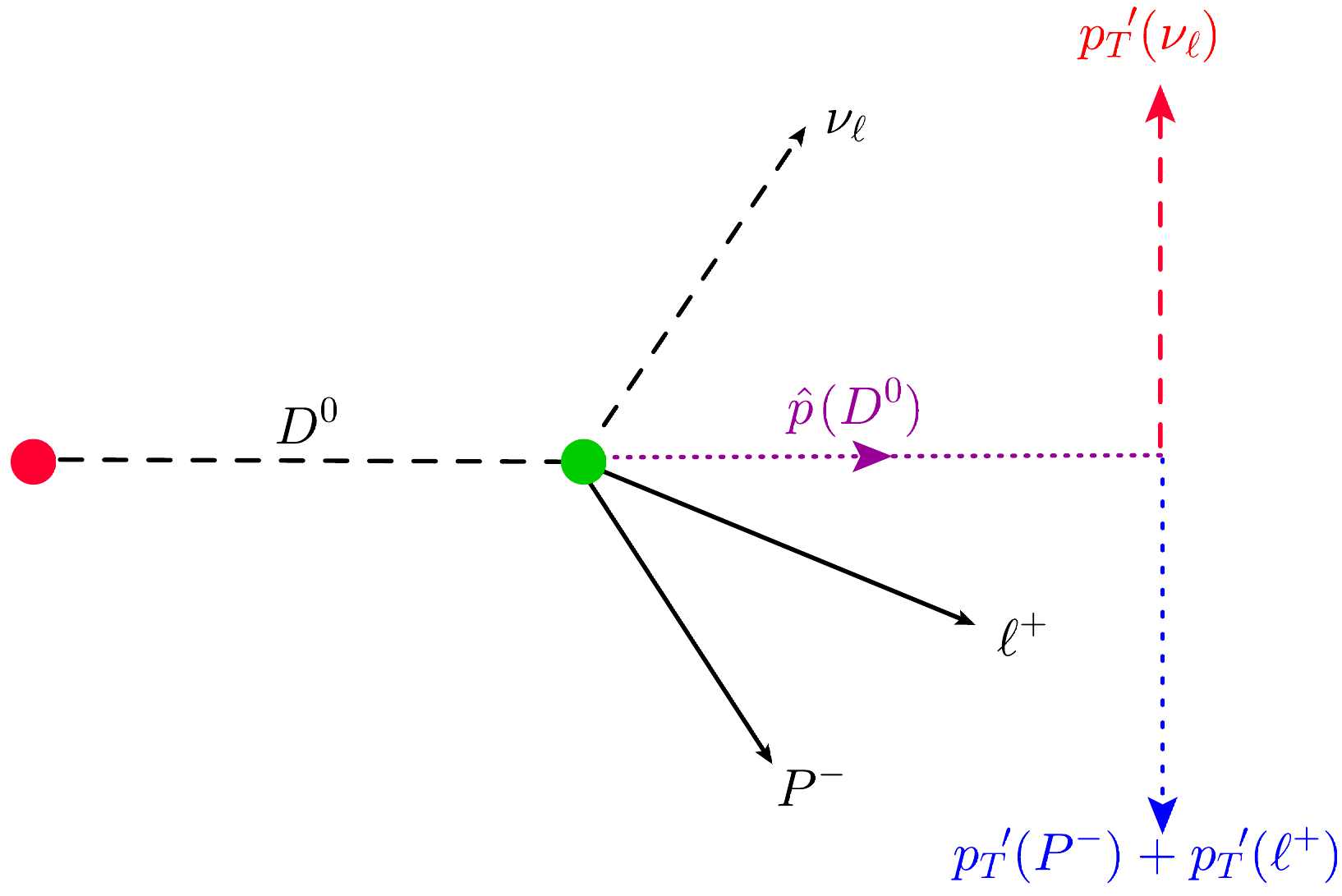}
\caption{Cartoon of the $D^0$ meson semileptonic decay, with the definition of the momentum component $p_T'(\nu)$. The red and green dots represent the origin and decay vertex of the $D^0$, respectively, the arrowed black lines represent the momentum components of the final state particles, the purple line indicates the initial flight direction of the $D^0$, and the red and blue
dashed lines represent the component of the momentum $p_T'$ perpendicular to the $D^0$ flight direction.}
\label{fig:ptprime}
\end{center}
\end{figure}

One of the major challenges of measuring semileptonic decays at LHCb is the partially reconstructed final state. 
Unlike at $e^+e^-$ machines, the hadron collider environment 
does not allow reconstruction of the missing neutrino using beam energy constraints, nor separation 
of the decays of interest into hemispheres. However, a host of experimental techniques exist for 
reconstruction of the neutrino momentum and the $q^2$ distributions. From kinematics alone, the neutrino
momentum is completely constrained perpendicular to the $D^0$ flight direction. We label this transverse momentum
$p_T'(\nu) = -(p_T'(P)+p_T'(\ell))$. A sketch is given in Figure~\ref{fig:ptprime}.

Using energy and momentum conservation to solve for the remaining component of the 
neutrino momentum leaves a two-fold ambiguity. While possible, simply choosing one of
the solutions of the neutrino momentum can lead to biases in the distributions of interest,
specifically with respect to $q^2$. There are several well-known methods for calculating the missing neutrino momentum component, relying on a variety of different techniques:
\begin{itemize}
	\item $k$-factor method: Using simulated events, determine the factor $k$ as a function of visible daughter mass which predicts the true $D^0$ momentum. Such an approach usually determines an average value $\langle k \rangle$ written as $p(D^0) = \frac{p(K\ell)}{\langle k(m(K\ell))\rangle}$. This approach has been used previously in many analyses, one such example being the measurement of the $B^0$ meson oscillation frequency $\Delta m_d$\cite{Aaij:2016fdk}.
	\item Corrected $D^0$ mass: One can use as an approximation to the true $D^0$ mass the quantity $m_\text{Corrected}=\sqrt{m^2(K\ell)+|p_T'|^2}+p_T'$. This quantity has the usefulness that it will peak at the nominal mother mass for true decays and have a long tail that extends to lower $m_\text{corrected}$. Decays of multibody final states will then peak more strongly towards lower $m_\text{corrected}$ allowing for good separation between signal and background. Such a method has been used successfully in the measurement of $|V_{ub}|$ from $\Lambda_b\to p\mu\nu$ decays \cite{Aaij:2015bfa}.
	\item Cone-closure: By enforcing that the $D^0$ be the daughter of a $D^{*+}\to D^0 \pi^+$ decay, the additional mass constraint of the $D^{*+}$ breaks the ambiguity of the momentum of the neutrino. The method relies on the fact that in the $K\ell$ rest frame, $p(D^0) = p(\nu)$, and the slow pion forms the axis of a cone around which the neutrino momentum lies. The solution of the momentum is then the one that most closely aligns the $D^0$ momentum to the $D^0$ flight direction. Such a procedure was used in the E687 experiment at Fermilab \cite{Johns:1995jc}.
	\item Recently, a new method using multivariate regression to help choose the correct neutrino momentum of $b$-hadron has been presented \cite{Ciezarek:2016lqu}. While the use case presented is for $b$-hadron decays, the algorithm should be easily extendible to $c$-hadron decays.
\end{itemize}

Each of these methods has its shortcomings, but provides an estimate of the momentum of the $D^0$ and thus the
calculation of $q^2$.

\section{Sensitivity for CKM matrix element measurements}
As an example of an accessible measurement at LHCb, we estimate the sensitivity of the measurement of the ratio of CKM 
matrix elements $|V_{cs}|/|V_{cd}|$. The measurement can be made by comparing the branching ratios of $D^{*+}\to D^0\pi$ 
with $D^0\to K^- \mu^+ \nu_mu$ and $D^0\to \pi^- \mu^+ \nu_\mu$. Such a measurement is tractable at LHCb due to the high 
efficiency of reconstructing muons. By enforcing that the $D^0$ originate from a $D^{*+}$ decay, the $q^2$ dependence of
both channels can be made unambiguously. This is a very similar measurement to the measurement in \cite{Aaij:2015bfa}.
The advantage of such a ratio measurement is that the majority of the trigger, selection and detection efficiencies cancel in the 
ratio. Additionally, the corrected mass can provide a stable handle on missing neutral backgrounds.

The sensitivity estimate can be made in the following way: as the branching ratio of the decay $D^\to K^-\mu^+\nu_\mu$ is
similar to that of $D^0\to K^-\pi^+$, an initial estimate of yields can be made for the Run I dataset by taking number of $D^0\to K^-\pi^+$ decays from the CPV search in charm using the same decay \cite{Aaij:2013wda}. This gives an initial estimate of $\sim$56 million $D^0\to K^-\mu^+\nu_\mu$ decays, and roughly one order of magnitude fewer decays of $D^0\to\pi^-\mu^+\nu_\mu$ decays. The software trigger to use for such a decay at the LHCb experiment is one which searches for inclusive $D^{*+}\to D^0 \pi^+$ decays. Such a trigger was only operational during two-thirds of the 2012 Run I operation, limiting the statistics
to roughly 60\% of the nominal value. Assuming that the remaining efficiency differences are at the order of 20\%, as the major
differences will be reconstructing the muon and forming a good vertex, this leaves roughly $4.4$ million signal candidates.
This yield would directly relate to a relative systematic uncertainty of 0.2\%.

\begin{figure}[htbp]
\begin{center}
\includegraphics[width=0.7\textwidth]{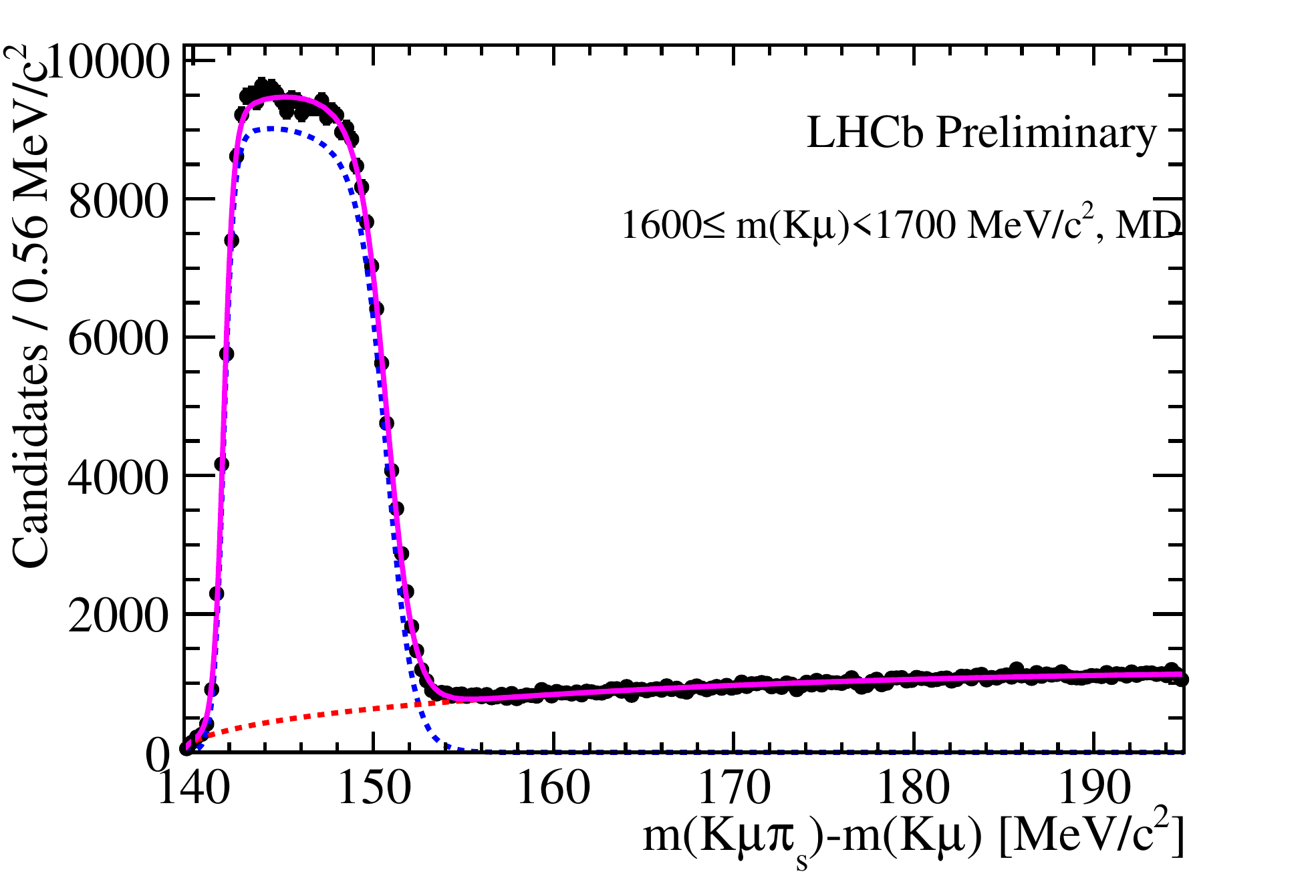}
\caption{Fit to the visible difference in mass $m(K\mu\pi)-m(K\mu)$ of the decay $D^0\to K\mu\nu$. The points correspond
to the sample of $D^0\to K\mu\nu$ decays in the 2012 magnet down dataset lying in the mass range $1600 \leq m(K\mu)<1700$ MeV/$c^2$. The total fit is shown in magenta, with the signal component shown in dashed blue, and the 
combinatorial background in dashed red.}
\label{fig2}
\end{center}
\end{figure}

These numbers are not simply out of thin air. A validation of the tracking efficiency in the analysis of $a_{sl}^s$\cite{Aaij:2016yze} used
the sample $D^0\to K^-\mu^+\nu_\mu$ to calculate the tracking asymmetry of the muon-pion pair. With loose selections, the analysis reconstructed $5M$ signal candidates. The total fit was performed simultaneously between $D^0\to K^- \mu^+\nu_\mu$ and $\overline{D}^0\to K^+ \mu^-\overline{\nu}_\mu$ samples in individual bins of visible mass $m(K\mu)$, with the signal shape being derived empirically. An example plot of the visible mass difference $m(K\mu\pi)-m(K\mu)$ for the 2012 dataset using the magnet down polarity fit in the range of visible mass $1600 \leq m(K\mu)<1700$ MeV/$c^2$ is shown in Figure~\ref{fig2}.

\section{Lepton non-universality in semileptonic $D$ decays}
\begin{figure}[htbp]
\begin{center}
\includegraphics[width=0.7\textwidth]{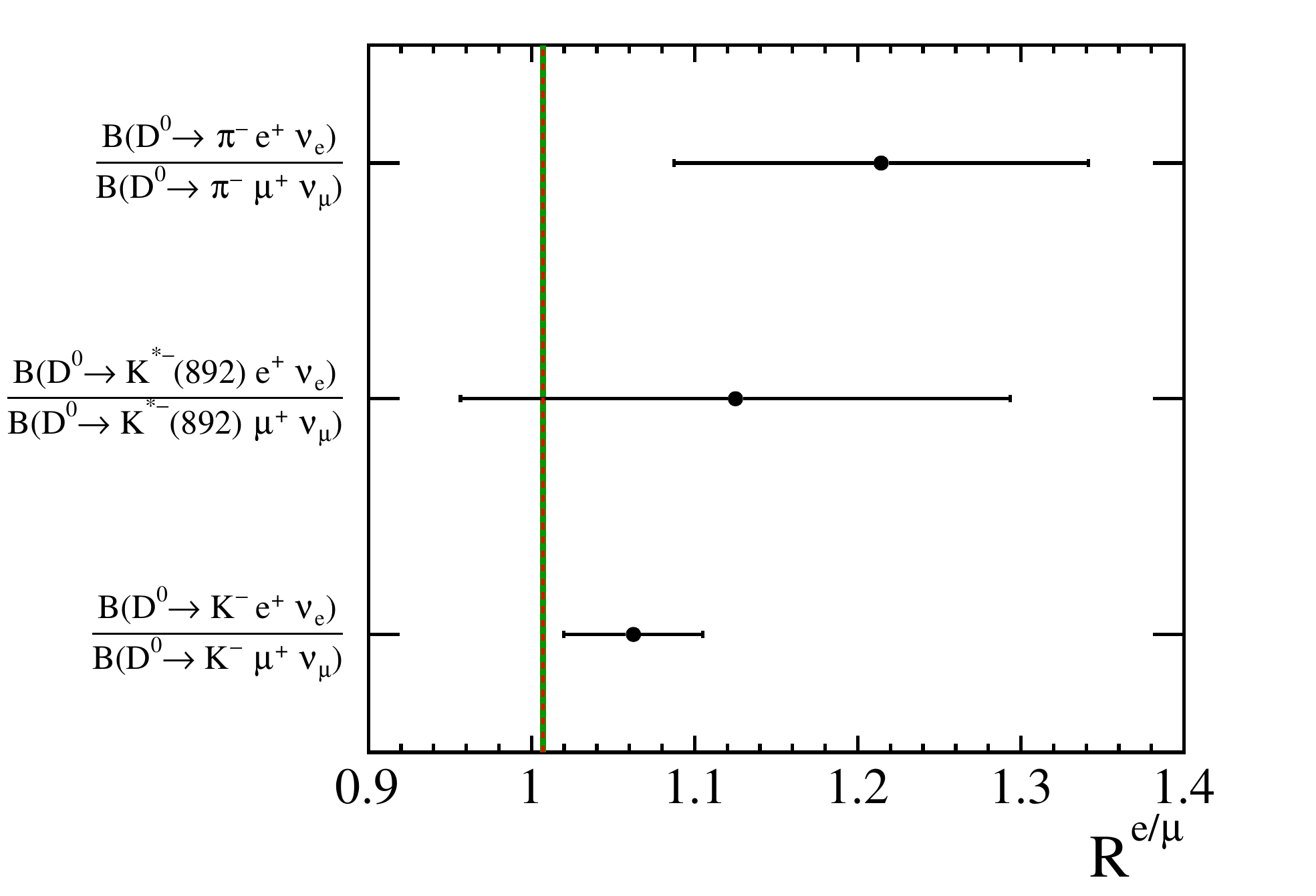}
\caption{Comparison of the ratio of measured branching fractions for $D^0\to h^- e^+\nu$ and $D^0\to h^-\mu^+\nu$. The term $h$ denotes the hadron in question, either $K^-$, $\pi^-$ or $K^{*-}(892)$. Each point represents the ratio from the measurements reported in \cite{Olive:2016xmw} compared to the standard mode prediction (red dashed line) and its error(green band) provided by S. Fajfer using~\cite{Koponen:2013tua}. The central values of each
point lie systematically to one side of the standard model prediction.
}
\label{fig3}
\end{center}
\end{figure}

Lepton non-universality has been a hot topic in the past few years mainly driven by possible anomalies in the semileptonic
decays of $B\to D^*\ell\nu$ with a $\tau$ in the final state compared to a $\mu$, as well as possible differences seen in decays of the form $b\to s \ell\ell$. Such tensions with the standard model beg the question whether or not there is a similar
measurement to be made in other modes. No such measurement of lepton non-universality has been made in the charm sector,
but each of the individual relevant branching ratios has been measured \cite{Olive:2016xmw}. By taking the simple ratio of relevant decays, and comparing with the $q^2$ integrated standard model prediction from~\cite{Fajfer:2015ixa} shows a consistent trend towards higher ratios than expected. This is illustrated in Figure~\ref{fig3}. 

All of the modes presented in Figure~\ref{fig3} are worth pursuing. Recent theory activity~\cite{Fajfer:2015ixa} shows that the measurement of the ratio $\mathcal{B}(D^0\to K^-e^+\nu_e) / \mathcal{B}(D^0\to K^-\mu^+\nu_\mu)$, for example, specifically as a function of $q^2$, allows for a direct probe of current bounds on allowed scalar Wilson coefficients. By using the same statistics quoted above for $D^0\to K^-\ell^+\nu_\ell$, the measurement of the ratio of the electron to muon modes would reduce the error on the bottom point in Figure~\ref{fig3} by an order of magnitude.
While the measurement of the electron mode is certainly difficult at LHCb, due to the fact that bremsstrahlung recovery is not easy, and the measurement of the tracking efficiencies of
the electron candidate is non-trivial, the measurement is not impossible; LHCb has already measured many channels with electrons, including the angular analysis of $B^0\to K^{*0}e^+e^-$\cite{Aaij:2015dea} and the search for lepton flavor violation in the decay of $D^0\to e^\pm \mu^\mp$\cite{Aaij:2015qmj}. 
\section{Conclusion}
We present preliminary estimates on the reach of LHCb in the field of semileptonic $D$ meson decay. We find that a measurement of $|V_{cs}|/|V_{cd}|$ would give a relative statistical uncertainty of $\sim$0.2\% using the Run I dataset. We also motivate the first search for lepton non-universality in the charm sector. It is important to note that all estimates are using the Run I dataset, and LHCb continues to take its Run II dataset with more statistics and improved triggering strategies. 

\end{document}